\documentclass[12pt]{iopart}
\usepackage{graphicx}
\usepackage{graphics}
\usepackage{color}
\usepackage{hyperref}
\usepackage[
    backend=biber,
    natbib=true,
    url=false, 
    doi=true,
    eprint=false
]{biblatex}
\begin{document}

\title{Static field limit of excitation probabilities in laser-atom interactions }

\author{A. Galstyan$^1$, V.L. Shablov$^2$, Yu.V. Popov$^{3,4}$, F. Mota-Furtado$^5$, P.F. O'Mahony$^5$,
              and B. Piraux$^1$}

\address{
$^1$Institute of Condensed Matter and Nanosciences, Universit\'e Catholique de Louvain,\\
         Chemin du Cyclotron 2/L7.01.07, B-1348 Louvain-la-Neuve, Belgium\\
$^2$Obninsk Institute for Nuclear Power Engineering of the
         National Research Nuclear University MEPhI, Obninsk, Russia\\
$^3$Skobeltsyn Institute of Nuclear Physics, Moscow State University, Moscow, Russia\\
$^4$Joint Institute for Nuclear Research, Dubna, Russia\\
$^5$Department of Maths, Royal Holloway, University of London, Egham, Surrey TW20 0EX, United Kingdom }

\begin{abstract}

We consider the interaction of atomic hydrogen, in its ground state, with an electromagnetic pulse whose duration is fixed in terms of the number of optical cycles. We study the probability of excitation of the atom in the static field limit {\it i.e.} for field frequencies going to zero. Despite the fact that the well known Born-Fock adiabatic theorem is valid only for a system whose energy spectrum is discrete, we show that  it is still possible to use this theorem to derive, in the low frequency limit, an analytical formula which gives the probability of transition to any excited state of the atom as a function of the field intensity, the carrier envelope phase and the number of optical cycles within the pulse. The results for the probability of excitation to low-lying excited states, obtained with this formula, agree with those we get  by solving the time dependent  Schr\"odinger equation. The domain of validity is discussed in detail.

\end{abstract}


\maketitle

\section{INTRODUCTION}

Experimental studies of  the interaction of atoms with strong near infrared laser pulses have shown that, in the tunnel ionization regime,  atomic excited states play a significant role. Nubbemeyer {\it et al.} have shown that, in the case of helium and for a broad range of field intensities up to $10^{15}$ Watt/cm$^2$, a substantial fraction of  atoms survive the laser pulse in many excited states \cite{nubbemeyer08,zimmermann17}. Their experimental data suggest that the excited state population trapping is predominantly due to a recombination process that they called frustrated tunneling. More recently, Beaulieu {\it et al.} studied the spectral, spatial and temporal characteristics of the radiation produced near the ionization threshold of Ar by few-cycle laser pulses \cite{Beaulieu16}. They showed that low-lying excited states are populated through multiphotonic absorption thereby leading to either direct extreme ultraviolet emission through free induction decay or to high-order harmonic generation through ionization from these low-lying excited states and recombination to the ground state.\\

Whether excited states are populated through frustrated tunneling or multiphoton transitions is still a matter of debate  \cite{lia14,lib14,piraux17}. As stressed by Morishita {\it et al.} \cite{Morishita13}, the division of strong field ionization of atoms by low-frequency infrared lasers in terms of multiphoton and tunneling ionization is too simple. However, since tunneling is actually a static concept, it makes sense to study the excitation probability in the quasi-static field limit. This is the objective of the present paper. Several attempts have been made to study this problem by solving numerically the time dependent Schr\"odinger equation  (TDSE) \cite{lia14,lib14,piraux17}. However, 
such a calculation is tremendously difficult and still out of reach in the limit where the frequency tends to zero.\\

Here, we consider the case of atomic hydrogen initially in its ground state and we calculate analytically the behavior of the excitation probabilities in the static field limit, which corresponds to the laser frequency becoming increasingly small while keeping constant the number {\it N} of optical cycles within the pulse. To do this we write the hamiltonian and the Schr\"odinger equation in scaled  time, and transform the hamiltonian by using complex scaling. In the adiabatic limit, the scaled time becomes a parameter and the hamiltonian reduces to the Stark hamiltonian. Herbst and Simon \cite{Herbst2} have shown that  the spectrum of the complex scaled Stark hamiltonian is purely discrete.  Under these conditions, it is possible to use the well-known Born-Fock adiabatic theorem to derive an analytical formula for the probability of excitation of the atomic ground state to any excited state in the limit where the carrier frequency goes to zero.\\

Our results provide  valuable information on the parameters that govern the electron dynamics at very low frequency. We study how the analytic formula for the excitation probability depends on the frequency, intensity, number of cycles and the Carrier Envelope Phase (CEP). Whenever it is possible, we compare our results with the excitation probabilities calculated by solving numerically the TDSE. Although very sophisticated methods have been developed \cite{tol1,tol2,tol3,baekhoj}, it is not always possible to reach numerically this static field limit. This comparison shows that the analytical formula describes accurately the excitation of low-lying excited states. 
Furthermore, for short pulse durations, it remains valid at high peak intensities where the interaction process is highly non perturbative. As far as the excitation to high-lying excited states is concerned, the analytical formula is not valid unless the frequency is extremely small. In fact, for any small frequency, it is always possible to find Rydberg excited states that will be resonantly coupled. This problem is discussed in detail.\\

The present contribution is organized as follows. After defining the basic equations, we derive an exact analytical formula that gives the probability for atomic hydrogen initially in its ground state, to be excited to a given bound state. We consider in detail the excitation of hydrogen to the 2s and 2p atomic states. By comparing our results with those obtained by solving numerically the TDSE, we analyze and discuss the validity of this analytical formula as a function of the frequency, the peak intensity, the pulse duration and the CEP before concluding.\\

Atomic units are used throughout unless otherwise specified.

\section{BASIC EQUATIONS}

We consider the interaction of atomic hydrogen initially in its ground state with a linearly polarized laser pulse. We work within the dipole approximation and describe the laser field in terms of a vector potential  given by
\begin{equation}
\vec{A}(t)=A_0f(t)\sin(\omega t+\phi)\;\vec{e}.
\end{equation}
$\vec{e}$ is a unit vector defining the polarization direction, $\phi$ is the CEP, $\omega$ is the field frequency,
$A_0$ is the amplitude of the vector potential and $f(t)$ is the pulse envelope.  By using the usual relation ($\vec{E}=-d\vec{A}/dt$) between the electric field $\vec{E}$ and the vector potential $\vec{A}$, it is easy to show that the amplitude of the electric field $E_0=\omega A_0$. We assume that $0 < t < T_N$ where $T_N$ is the duration of the pulse given by $T_{N}=NT=N(2\pi/\omega)$ where $N$ is the total number of optical periods $T$ within the pulse. In addition, we impose $f(0)=f(NT)=0$.  In the present context, it is convenient to introduce the scaled time $\tau=\omega t$. In terms of $\tau$,  the duration of the pulse is given by $\tau_N=2\pi N$. \\

The total hamiltonian $H(\tau)=H_0+V(\tau)$, in terms of the scaled time, is the sum of the atomic hamiltonian $H_0$ for the hydrogen atom and the dipole interaction operator $V(\tau)$, in its Length (L) gauge form. In the configuration space, we have:
\begin{eqnarray}
H_0&\equiv&-\frac{1}{2}\triangle-\frac{1}{r},\\
V(\tau)&\equiv&\vec{r}\cdot\vec{E}(\tau),
\end{eqnarray}
where the electric field $\vec{E}(\tau)$ is obtained from the  relation $\vec{E}(\tau)=-\omega\; d\vec{A}/d\tau$.
The TDSE is
\begin{equation}
\label{eq_tdse_full}
\left[\mathrm{i}\omega\frac{\partial}{\partial\tau}+\frac12\triangle+\frac{1}{r}-(\vec E(\tau)\cdot\vec
r)\right]{\tilde\Phi^{(L)}}(\vec r,\tau)=0,
\end{equation}
with the initial condition
\begin{equation}
\tilde{\Phi}^{(L)}(\vec{r},0)=\varphi_{1s}(\vec{r})=\frac{1}{\sqrt{\pi}}e^{-r}.
\end{equation}
The tilde above the wave function indicates the fact that it is expressed in the configuration space. This wave function is normalized in the usual way
\begin{equation}
\int |{\tilde\Phi^{(L)}}(\vec r,\tau)|^2\ \mathrm{d}^3r=1.
\end{equation}
We want to find an analytical formula, which allows for the calculation of the low frequency behavior of the wave function
 $\tilde{\Phi}^{(L)}(\vec{r},\tau)$. There are however two observations, which make that task difficult. Firstly, for $\omega\to 0$, Eq. (\ref{eq_tdse_full}) belongs to a class of equations called "singularly perturbed differential equations". These differential equations contain a small parameter in front of the highest derivative. Many rigorous mathematical studies of this problem have been performed and have led to well known results \cite{VB1, VB2}. However, none of them can be used here because of the presence, in Eq. (\ref{eq_tdse_full}), of the imaginary unit. Secondly, in the limiting case where the electric field becomes time independent, the total hamiltonian $H(\tau)$ becomes $\tau$ independent and coincides with the so-called Stark hamiltonian for atomic hydrogen. This hamiltonian has a  complicated spectrum which has been the subject of much research in mathematics 
\cite{ReedSimon,titsch}. These two remarks clearly show that the mathematical treatment of the present problem is very difficult. \\

In order to calculate the probability of excitation from the ground state to a given excited state, we use the time evolution operator $U(\tau)$ which, in Dirac notation, is such that
\begin{equation}
\label{eq_evolution}
|{\Phi^{(L)}}(\tau)\rangle=U(\tau)|{\Phi^{(L)}}(0)\rangle,
\end{equation}
and satisfies the differential equation
\begin{equation}
\mathrm{i}\omega\frac{\partial U(\tau)}{\partial \tau}=H(\tau)U(\tau),
\end{equation}
with the initial condition $U(0)=I$. The probability of transition from the ground state to any excited state $|\varphi\rangle$ of the hydrogen atom is defined at time $\tau = \tau_N$ as
\begin{equation}
\label{eq_prob_trans}
W_{s0}=|\langle\varphi_s|U(\tau_N)|{\Phi^{(L)}}(0)\rangle|^2.
\end{equation}
This is the quantity we want to calculate to find the probability of excitation to a given state in the static field limit.

\section{LOW FREQUENCY LIMIT OF THE EXCITATION PROBABILITY}
\subsection{Preliminary remarks}
In taking the limit $\omega \rightarrow 0$, we need to deal with the spectrum of the Stark hamiltonian for atomic
hydrogen. It is well known that the field free atom has a discrete bound state spectrum starting from a lowest finite energy and a continuous spectrum starting at the continuum threshold. In contrast, with an arbitrarily small electric field, the spectrum becomes purely continuous, ranging from $-\infty$ to $+\infty$. The bound state energies of the field free atom become complex with a negative imaginary part or in other words, turn into resonances \cite{ReedSimon,titsch,Herbst1a,Herbst1b,Herbst2}. This result can be shown by means of a complex scaling of the hamiltonian \cite{ReedSimon,titsch,Simon}. In the configuration space, the electron radial coordinate is scaled according to $r\rightarrow re^{\theta}$ where $\theta$ is complex and usually purely imaginary. In the present case, we perform such complex scaling of the hamiltonian $H(\tau)$. Under  these conditions, Eq. (\ref{eq_evolution}) and (\ref{eq_prob_trans}) become
\begin{eqnarray}
|{\Phi^{(L)}}(\tau,\theta)\rangle&=&U(\tau,\theta)|{\Phi^{(L)}}(0,\theta)\rangle,\\
\mathrm{i}\omega\frac{\partial U(\tau,\theta)}{\partial \tau}&=&H(\tau,\theta)U(\tau,\theta).
\end{eqnarray} 
For  $0< |Im\ \theta|<\pi/3$, Herbst and Simon \cite{Herbst2} have shown that the spectrum of the hamiltonian $H(\tau,\theta)$ is purely discrete. This result is very important because it allows us to use, in the following, a treatment based on the well known Born-Fock adiabatic theorem \cite{BornFock}. However, we are no longer working in a Hilbert space. It is therefore necessary to properly define the inner product and a closure relation. We have 
\begin{eqnarray}
H(\tau,\theta)|\psi_j(\tau,\theta)\rangle&=&\varepsilon_j(\tau)|\psi_j(\tau,\theta)\rangle,\\
\langle\psi_j(\tau,\theta^*)|H(\tau,\theta)&=&\varepsilon_j(\tau)\langle\psi_j(\tau,\theta^*)|.
\end{eqnarray}
The discrete index $j$ runs over all possible eigenstates including degenerate eigenstates. It is important to note that the eigenvalue $\varepsilon_j(\tau)$ does not depend on $\theta$ \cite{Herbst2}. The eigenstates $|\psi_j(\tau,\theta)>$ are normalized according to
\begin{equation}
\label{eq_norm}
\langle\psi_j(\tau,\theta^*)|\psi_{j'}(\tau,\theta)\rangle=\delta_{jj'}.
\end{equation}
If we define the projector operator
\begin{equation}
P_j(\tau,\theta)=|\psi_{j}(\tau,\theta)\rangle\langle\psi_j(\tau,\theta^*)|,
\end{equation}
the closure relation can be written as follows
\begin{equation}
\label{eq_closure}
\sum_j\  P_j(\tau,\theta)=I.
\end{equation}
In the case of degenerate states, Eq. (\ref{eq_closure}) includes an additional summation over indices corresponding to such states. Proof of this closure relation is analogous to the proof given in \cite{Simon} for a N-particle system.\\

Since the set of eigenvectors $|\psi_{j'}(\tau,\theta)\rangle$ is complete, we decompose our wave packet $|{\Phi^{(L)}}(\tau,\theta)\rangle$
as follows
\begin{equation}
\label{eq_decompose}
|{\Phi^{(L)}}(\tau,\theta)\rangle=U(\tau,\theta)|{\Phi^{(L)}}(0,\theta)\rangle=\sum_j C_j(\tau,\theta)|\psi_{j}(\tau,\theta)\rangle,
\end{equation}
with
\begin{equation}
C_j(\tau,\theta)=\langle\psi_j(\tau,\theta^*)|U(\tau,\theta)|{\Phi^{(L)}}(0,\theta)\rangle.
\end{equation}
After inserting Eq. (\ref{eq_decompose}) into the TDSE, using the normalization condition (\ref{eq_norm}) and defining
\begin{eqnarray}
\label{19}
f_{sj}(\tau,\theta)&=&\left\langle\psi_s(\tau,\theta^*)\left|\frac{\partial\psi_j(\tau,\theta)}{\partial\tau}\right.\right\rangle \nonumber\\
&= &- \frac{\langle\psi_s(\tau,\theta^*)|\partial V(\tau,\theta)/\partial\tau|
\psi_{j}(\tau,\theta)\rangle}{[\varepsilon_s(\tau)-\varepsilon_j(\tau)]}, \quad s\neq j,
\end{eqnarray}
we obtain the following system of equations for the coefficients $C_j(\tau,\theta)$,
\begin{eqnarray}
\label{eq_tdse_rewritten2}
\mathrm{i}\omega \frac{\partial C_s(\tau,\theta)}{\partial\tau}&=&\left[\varepsilon_s(\tau)-\mathrm{i}\omega\langle
\psi_s(\tau,\theta^*)|\frac{\partial\psi_{s}(\tau,\theta)}{\partial\tau}\rangle\right] C_s(\tau,\theta)\nonumber\\
&-&\mathrm{i}\omega\sum_{j\neq s}\ f_{sj}(\tau,\theta) C_j(\tau,\theta),
\end{eqnarray}
with $ C_s(0,\theta)=\delta_{s0}$.  It follows from \cite{Messiah} that the second term in square brackets in Eq. (\ref{eq_tdse_rewritten2}) vanishes (see also the Appendix). \\ 

At this stage, it is important to stress that in the Born-Fock adiabatic theorem, which is presented in many textbooks on quantum mechanics (see, for example, \cite{Messiah,Schiff,Joachain}), it is tacitly assumed that the hamiltonian $H(\tau)$ has only a discrete spectrum. This rather abstract situation happens only in the case of an harmonic  potential. However, the above discussion shows that it is also true in the case of the complex scaled  Stark hamiltonian $H(\tau,\theta)$; such a hamiltonian is no longer self-adjoint but has a discrete spectrum originating from the discrete Coulomb spectrum if $0< |Im\ \theta|<\pi/3$. Let us now discuss in more detail relation (\ref{19}).  On the one hand, it was proven in \cite{BornFock} that it is always possible to find a constant $M_{sj}$ independent on $\tau$ and such that $|f_{sj}(\tau,\theta)|<M_{sj}$. On the other hand, the equality
$$
\left\langle\psi_s(\tau,\theta^*)\left|\frac{\partial\psi_j(\tau,\theta)}{\partial\tau}\right.\right\rangle = - \frac{\langle\psi_s(\tau,\theta^*)|\partial V(\tau,\theta)/\partial\tau|\psi_{j}(\tau,\theta)\rangle}{[\varepsilon_s(\tau)-\varepsilon_j(\tau)]}
$$
is valid if $s\neq j$. As it is  shown in \cite {Kato}, the difference $[\varepsilon_s(\tau)-\varepsilon_j(\tau)]$ never equals zero except in the case of degenerate states for times $\tau_q$ where $E(\tau_q)=0$. If so, it is crucial that
$$
E'(\tau_q)\langle\psi_s(\tau_q,\theta^*)|\vec e\cdot\vec r|\psi_{j}(\tau_q,\theta)\rangle=0.
$$
If we suppose that $E'(\tau_q)\neq 0$, which is generally the case during the interaction with the pulse, the dipole matrix element $\langle\psi_s(\tau_q,\theta^*)|\vec e\cdot\vec r|\psi_{j}(\tau_q,\theta)\rangle$ must be equal to zero. This is not necessary the case for states $|\psi_j\rangle$ that are eigenstates of the angular momentum operator like for instance the $|2s\rangle$ and $|2p\rangle$ states. However, it can be shown that this dipole matrix element will be equal to zero if  $|\psi_j\rangle$ and $|\psi_s\rangle$ are degenerate parabolic states. Nevertheless, the difference $[\varepsilon_s(\tau)-\varepsilon_j(\tau)]$ can be rather small in some domains of values of $\tau$ and field intensities or in the case of high-lying Rydberg  states. In fact, to our knowledge, the value of the majorant $M_{sj}$ has never been estimated. This interesting problem needs  further investigation. \\

In many textbooks that discuss the Born-Fock adiabatic theorem, it is usual to write
$$
C_s(\tau, \theta) = a_s(\tau, \theta)\ \exp\left(-\frac{\imath}{\omega}\int_0^\tau
\varepsilon_s(\xi)d\xi\right), \quad a_s(0, \theta)\ =\delta_{s0}.
$$
It is of course always possible to proceed in this way. However, we have shown that in this case,  the coefficients $a_s$ are of exponential type. Instead, it is more convenient to suppose the following form of the coefficients $C_s(\tau, \theta)$ 
\begin{equation}
\label{form}
C_s(\tau,\theta)=\sum_{p=0}^{\infty}\
a_{sp}(\tau,\theta)\exp\left(-\frac{\imath}{\omega}\int_0^\tau
\varepsilon_p(\xi)d\xi\right), \quad \sum_{p=0}^{\infty}\ a_{sp}(0,\theta)=
\delta_{s0}.
\end{equation}
In this case, it is possible to show that we can find a Taylor series in power of $\omega$  for the coefficients $a_{sp}(\tau, \theta)$. Inserting Eq.(\ref{form}) into Eq.(\ref{eq_tdse_rewritten2}) we obtain 
\begin{eqnarray}
\label{sum exp}
\sum_{p=0}^{\infty}[\frac{\partial a_{sp}(\tau,\theta)}{\partial
\tau}+\frac{\varepsilon_p(\tau)-\varepsilon_s(\tau)}{\imath\omega}a_{sp}(\tau,\theta)
&+&\sum_{j\neq s}^{\infty}\ f_{sj}(\tau,\theta)a_{jp}(\tau,\theta)]\nonumber\\
&\times&\exp\left(-\frac{\imath}{\omega}\int_0^\tau\varepsilon_p(\xi)d\xi\right)=0. 
\end{eqnarray}
Eq.(\ref{sum exp}) is of the type of a general Dirichlet series :
$$
\sum_p^{\infty} b_p(\tau)e^{-q\ \gamma_p(\tau)}=0. 
$$
We assume that $q$, which is inversely proportional to $\omega$, is a quite big parameter, that the complex functions $\gamma_p(\tau)$ are independent of $\omega$ and $\mathcal{R}e( \gamma_p(\tau))>0$. We
also suppose that the coefficients $b_p(\tau)$ are not of exponential type. In this case, it is possible to show \cite{Hardy} that $b_p(\tau)=0$, implying:
\begin{equation}
\label{eq coef}
\frac{\partial a_{sp}(\tau,\theta)}{\partial
\tau}+\frac{\varepsilon_p(\tau)-\varepsilon_s(\tau)}{\imath\omega}a_{sp}(\tau,\theta)
+ \sum_{j\neq s}\ f_{sj}(\tau,\theta)a_{jp}(\tau,\theta)=0.
\end{equation}
Rigorously speaking, this last result is correct provided that $\mathcal{R}e( \gamma_p(\tau))$ is a strictly increasing sequence of non-negative real numbers that tend to infinity for increasing $p$. In the present case, $\mathcal{R}e( \gamma_p(\tau))$ is the imaginary part of the complex energy $\varepsilon_p$ integrated over $\tau$. It is increasing with $p$ but it is not clear that it tends to infinity. Nevertheless, we assume here that it is true and show {\it a posteriori} that this formulation provides exactly the results we obtain by following the Born-Fock approach.

\subsection{Exact low frequency limit}
In the following,  we assume the frequency $\omega$ sufficiently small with respect to $|\varepsilon_p(\tau)-\varepsilon_s(\tau)|$ so that the following expansions are valid,
\begin{equation}
\label{23.1}
a_{ss}(\tau,\theta)= \delta_{s0}+ \sum_{k=1}\ (\imath\omega)^k\
a_{ss}^{(k)}(\tau,\theta), 
\end{equation}
and for $s\neq p$
\begin{equation}
\label{23.2}
a_{sp}(\tau,\theta)= \sum_{k=1}\ (\imath\omega)^k\
a_{sp}^{(k)}(\tau,\theta). 
\end{equation}
The initial condition is such that
\begin{equation}
\label{24}
\sum_{p=0}\ a_{sp}^{(k)}(0,\theta)=0, \quad k=1,2,3, ... . 
\end{equation}
By inserting these expansions  in (\ref{eq coef}), we obtain the following system of equations for the coefficients
\begin{equation}
\label{27.1}
\frac{\partial a_{ss}^{(k)}(\tau,\theta)}{\partial \tau} +
\sum_{j\neq s}\ f_{sj}(\tau,\theta)a_{js}^{(k)}(\tau,\theta)=0;
\end{equation}
and for $s\neq p$,
\begin{equation}
[\varepsilon_p(\tau)-\varepsilon_s(\tau)]a_{sp}^{(1)}(\tau,\theta)+f_{sp}(\tau,\theta)\delta_{p0}=0;
\label{27.2}
\end{equation}
\begin{equation}
\frac{\partial a_{sp}^{(k)}(\tau,\theta)}{\partial \tau} +
[\varepsilon_p(\tau)-\varepsilon_s(\tau)]a_{sp}^{(k+1)}(\tau,\theta)+
\sum_{j\neq s}\ f_{sj}(\tau,\theta)a_{jp}^{(k)}(\tau,\theta)=0.
\label{27.3}
\end{equation}
We can now generate all coefficients by iteration. It follows from (\ref{27.2}) that
\begin{equation}
a_{s0}^{(1)}(\tau,\theta)=-\frac{f_{s0}(\tau,\theta)}{\varepsilon_0(\tau)-\varepsilon_s(\tau)},
\quad a_{sp}^{(1)}(\tau,\theta)=0\;\;\mathrm{with}\;\; p>0,\;\;  s\neq p. \label{28}
\end{equation}
In turn, Eq. (\ref{27.1}) gives
\begin{equation}
a_{00}^{(1)}(\tau,\theta)=\sum_{j\neq 0}\ \int_0^\tau d\xi\
\frac{f_{0j}(\xi,\theta)f_{j0}(\xi,\theta)}{\varepsilon_0(\xi)-\varepsilon_j(\xi)},
\quad a_{ss}^{(1)}(\tau,\theta)=B_s,\;\; s>0, \label{29}
\end{equation}
where $B_s$ is a constant which, keeping in mind Eq. (26), is  given by
\begin{equation}
B_s=-\ a_{s0}^{(1)}(0,\theta)= \
\frac{f_{s0}(0,\theta)}{\varepsilon_0(0)-\varepsilon_s(0)}. \label
{30}
\end{equation}
In this way, all the coefficients at the first order are defined, and we can use Eq. (29) to generate the coefficients at the second order, etc. We clearly see, that they are not of an exponential type. Finally, the term  $C^{(1)}_0(\tau,\theta)$ is given up to the first order in $\omega$ by, 
\begin{equation}
C^{(1)}_0(\tau,\theta)= \left[1+ \imath\omega
\sum_{j\neq 0}\ \int_0^\tau d\xi\
\frac{f_{0j}(\xi,\theta)f_{j0}(\xi,\theta)}{\varepsilon_0(\tau)-\varepsilon_j(\tau)}  \right]\exp\left(-\frac{\mathrm{i}}{\omega}\int_0^\tau\varepsilon_0(\xi)\mathrm{d}\xi\right). \label{zero}
\end{equation}
We also obtain the  terms $C^{(1)}_{s>0}(\tau,\theta)$  as follows
\begin{eqnarray}
\label{s}
C^{(1)}_{s>0}(\tau,\theta)= \imath\omega[&-&\frac{f_{s0}(\tau,\theta)}{\varepsilon_0(\tau)-\varepsilon_s(\tau)}\exp\left(-\frac{\mathrm{i}}{\omega}\int_0^\tau\varepsilon_0(\xi)\mathrm{d}\xi\right)\nonumber\\
&+&\frac{f_{s0}(0,\theta)}{\varepsilon_0(0)-\varepsilon_s(0)}\exp\left(-\frac{\mathrm{i}}{\omega}\int_0^\tau\varepsilon_s(\xi)\mathrm{d}\xi\right)].
\end{eqnarray}
At the end of the pulse where $\tau=\tau_N$, we rotate back the contour to the real axis ($\theta=0$). Under this condition, $f_{s0}(\tau_N,0)=f_{s0}(0,0)$. At this stage, our results show that the  probability amplitude to stay in the ground state ($s=0$) is given at order zero by
\begin{equation}
C^{(0)}_0(\tau_N,0)=\exp\left[-\frac{\mathrm{i}}{\omega}\int_0^\tau\varepsilon_0(\xi)\mathrm{d}\xi\right].
\end{equation}
Since $\varepsilon_0(\xi)=Re\left[\varepsilon_0(\xi)\right]-i\Gamma_0(\xi)/2$ where the ground state width $\Gamma_0(\xi) >0$, the probability to stay in the ground state  is
\begin{equation}
\label{eq_w00}
W_{00}=|C_{0}^{(0)}(\tau_N,0)|^2\approx \exp\left[-\frac{1}{\omega}\int_0^{\tau_N} \Gamma_0(\xi)\mathrm{d}\xi\right], 
\end{equation}
which is a well known result. On the other hand, the probability amplitude  for excitation to a specific state $s\neq 0$, calculated at the first order,  is given by
\begin{eqnarray}
\label{eq_cs_gr_0}
C_{s>0}^{(1)}(\tau_N,0)\approx \mathrm{i}\omega E'(0)\frac{\langle\varphi_s|(\vec e\cdot\vec r|\varphi_0\rangle}{[\varepsilon_0(0)-\varepsilon_s(0)]^2}
&\times&(\exp\left[-\frac{\mathrm{i}}{\omega}\int_0^{\tau_N}\varepsilon_0(\xi)\mathrm{d}\xi\right]\nonumber\\
&-&\exp\left[-\frac{\mathrm{i}}{\omega}\int_0^{\tau_N}
\varepsilon_s(\xi)\mathrm{d}\xi\right]).
\end{eqnarray} 
In our calculations, we consider a sine square pulse envelope $f(\tau)=\sin^2(\tau/2N)$. In that case, we have:
\begin{equation}
\label{eq_field_der}
E'(0)=-\omega\left[\frac{\partial^2 A(\tau)}{\partial\tau^2}\right]_{\tau=0}=-\frac{A_0}{2N^2}\sin(\phi).
\end{equation}
This immediately shows that the CEP $\phi$ and the number $N$ of cycles within the pulse are key parameters, at least for the present shape of the pulse. When $\phi=0$, the first order expansion (\ref{eq_cs_gr_0}) is identically zero thereby requiring the calculation of the second order terms in $\omega$. \\

To calculate the excitation probability amplitude (37), we proceed as follows. Let us assume that we want to calculate the probability of excitation to a given $|n \ell m\rangle$ state of atomic hydrogen. Since, in presence of a static electric field, the angular momentum is not preserved ($\ell$ is not anymore a good quantum number), the complex energies present in the exponential factors of Eq. (\ref{eq_cs_gr_0}) can no longer be associated to a single atomic hydrogen state except for the ground state. To solve this problem,  we need to move to the basis of parabolic states $|n_1 n_2 m\rangle$ as defined in \cite{Landau} where the principal quantum number $n=n_1+n_2+|m|+1$. In absence of electric field, a given atomic hydrogen state $|n \ell m\rangle$ can be written as a linear combination of parabolic states $|n_1 n_2 m>$ as follows:
\begin{equation}
|n \ell m\rangle=\sum_{n_1=0}^{n-|m|-1}\langle n_1n_2m|n\ell m\rangle|n_1 n_2 m\rangle,
\end{equation}
where \cite{Belkic},
\begin{equation}
\langle n_1 n_2 m|n\ell m\rangle=(-1)^{n_1+(3|m|-m)/2}\mathcal{C}^{\ell m}_{\frac{n-1}{2}\;\frac{m+n_1-n_2}{2}\;\frac{n-1}{2}\;\frac{m+n_2-n_1}{2}}.
\end{equation}
In this equation, $\mathcal{C}^{j m}_{j_1m_1 j_2 m_2} $ is a usual Clebsch-Gordon coefficient \cite{Jahn}. Consequently, the probability amplitude for excitation to an atomic hydrogen state $|n \ell m\rangle$ is given by:
\begin{equation}
\label{eq_cnlm}
\tilde{C}_{n\ell m}(\tau_N,0)=\sum_{n_1=0}^{n-|m|-1}\langle n_1n_2m|n\ell m\rangle\; C_{n_1 n_2 m}(\tau_N,0).
\end{equation}
From now on, we use a tilde when $C$ refers to the amplitude of probability of excitation to an atomic state of well defined angular momentum and no tilde when $C$ refers to the amplitude of probability of excitation to a parabolic state, the expression of which is given by Eq. (37). The label $s$ in Eq. (37) therefore refers to individual parabolic states (for $\tau=0$).

\section{NUMERICAL ASPECTS}
Let us now discuss briefly how the complex energies that enter the calculation of $C_{n_1 n_2 m}(\tau_N,0)$ are obtained numerically. We calculate the complex rotated Stark hamiltonian in a basis of sturmian functions in parabolic coordinates\footnote{These functions are solution of the time independent radial Schr\"odinger equation in parabolic coordinates for atomic hydrogen. In this equation, the eigenenergy is fixed and negative while the eigenvalue is the nuclear charge. These sturmian functions form a complete and discrete set of basis functions.} and then diagonalize this hamiltonian. It is important to stress that in the Herbst and Simon theorem \cite{Herbst2}, the static electric field is present in the whole space. Numerically however, we use a finite basis or, equivalently, consider a finite box. As a result, in addition to the resonances, the spectrum of the complex rotated Stark hamiltonian contains two continua \cite{Cerjan}. To "follow" each complex energy as a function of the electric field strength, it is necessary to diagonalize the Stark hamiltonian many times and, for each field strength, to select, among all the eigenvalues (many of which corresponding to the continua), the resonance we are interested in. Since the size of our sturmian basis can be very large in the case of high field strengths, we use the well known Lanczos algorithm which allows one to calculate a small number of  complex eigenvalues around the one we are looking for and to keep the computer time within reasonable limits. This procedure works very well for low lying parabolic states. The eigenvalues we obtain are in excellent agreement with the most accurate ones available in the literature \cite{FFernandez}.\\

In order to assess the validity of expression (37) for the amplitude of the probability of excitation to a given state in the low frequency regime, we study, in the following Section,  the probability of excitation to 2s and 2p (m=0) states of atomic hydrogen initially in its ground state and subjected to a linearly polarized field. This probability is studied  as a function of the frequency for different peak intensities, pulse durations and CEPs. We compare our results obtained with Eq. (37) and by solving numerically the TDSE.\\

As briefly described in \cite{piraux17}, our numerical method to solve the TDSE is based on a spectral method, which consists in expanding the total wave function in a basis composed of products of spherical harmonics and complex radial Coulomb Sturmian functions. These Sturmian functions depend on a nonlinear parameter which allows one to monitor the region of the bound state spectrum we want to describe accurately. However, at very low frequencies, we expect the number of angular momenta needed to become increasingly large. In fact, high values of the angular momentum characterize the continuum states. By using complex Sturmian functions, equivalent to the use of real Sturmian functions with a global complex rotation of the total Hamiltonian, the continuum electron flux is rapidly absorbed  preventing the migration to very high values of the angular momentum. In all our calculations based on the numerical solution of the TDSE, convergent results are obtained by using 2000 Sturmian functions per angular momentum and 192 values of this angular momentum. Failing to use the global complex rotation of the total Hamiltonian gives results that do not converge in terms of the number of angular momenta for the lower frequencies considered here. \\

\section{RESULTS AND DISCUSSION}

Before studying the excitation to the $n=2$ states of atomic hydrogen, initially in its ground state, let us first start with the following preliminary remark. Quasi crossings of complex energies of different $n$-manifolds, occur for rather low electric fields \cite{Fernandez}. By quasi crossing of complex energies, we mean that $|\varepsilon_s(\tau)-\varepsilon_{s'}(\tau)|/\omega \leq 1$. In the case of the $n=2$-manifold, quasi crossings with energy curves belonging to the $n=3$-manifold  start to occur at intensities around 10$^{12}$ W/cm$^2$. Beyond this intensity, the parabolic states lose their identity. This is expected to reduce the domain of validity of the previous analytical formula to rather low peak intensities. However, for 10$^{12}$ W/cm$^2$, the imaginary part of the complex energies is very small so that the absolute value of the exponential terms in Eq. (37) remains, for short pulse durations, close to one thereby leading to a pure $\omega^2$ dependence for the probability of excitation to the $n=2$ states. For the excitation to high lying Rydberg states, the problem is similar but, the quasi crossings occur, as expected, at much lower intensities.\\

\begin{figure}[h!]
\includegraphics[width=13cm,height=8cm]{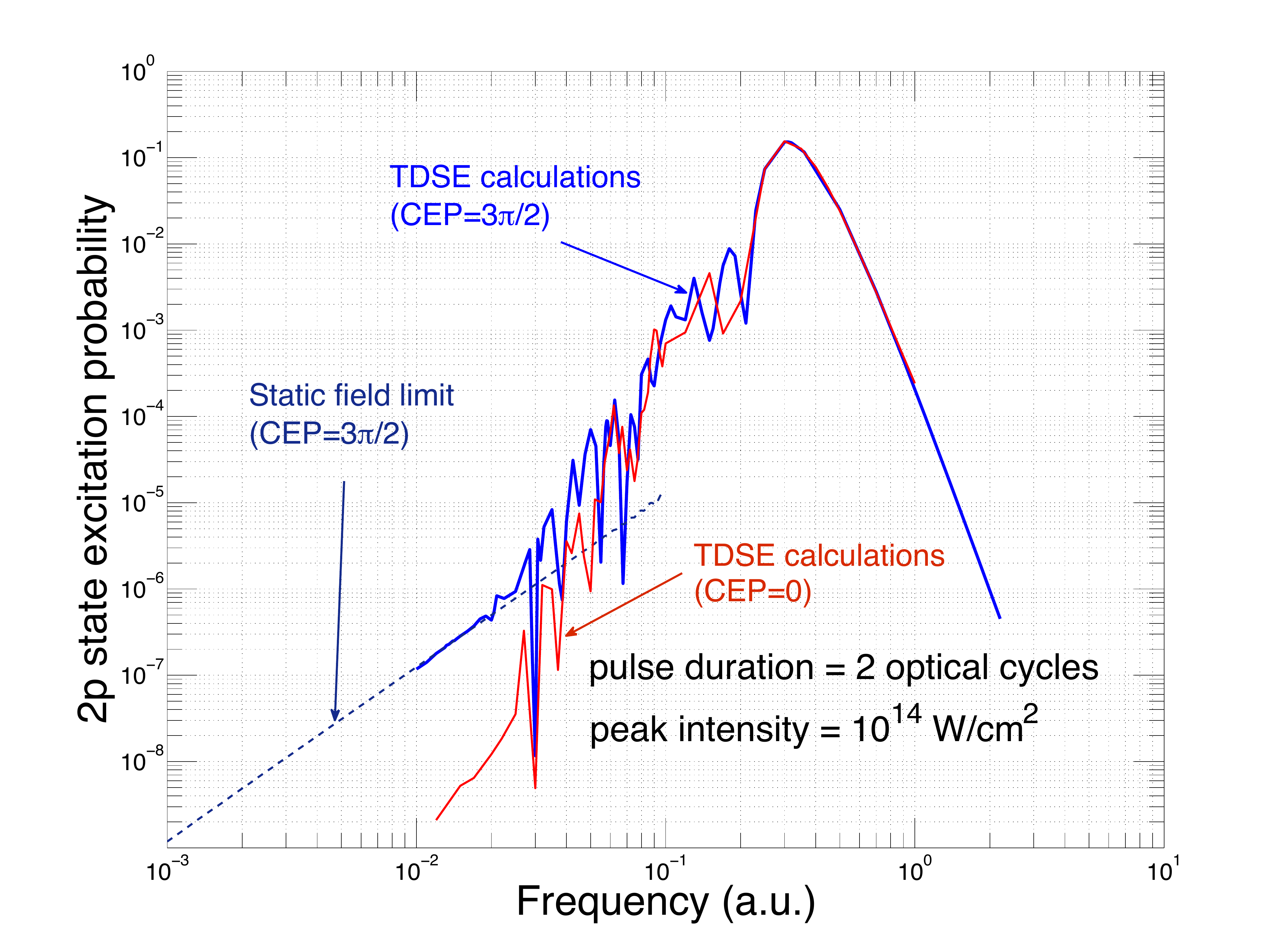}
\caption{(Color online) 2p state excitation probability as a function of the frequency in the case of the interaction of atomic hydrogen in its ground state with a 2 optical cycle sine square pulse of $10^{14}$ W/cm$^2$ peak intensity. Results obtained by solving numerically the TDSE are presented for two different values of the CEP: CEP=0, red full line and CEP=$3\pi/2$, blue full line. For CEP=$3\pi/2$, the TDSE results are compared to those obtained within the static field limit (Eq. 37). For CEP=0, this limit gives zero.}
\label{fig1}
\end{figure}

Let us first consider the interaction of atomic hydrogen in its ground state with a 2 optical cycle sine square pulse of $10^{14}$ W/cm$^2$ peak intensity. In Fig. \ref{fig1}, we study the probability of excitation to the 2p state as a function of the frequency. Results obtained by solving numerically the TDSE are presented for two values of the 
%%%%%%%%%%% FIGURE 1
%%%%%%%%%%%%%%%%% 
CEP, $\phi=0$ and $\phi=3\pi/2$. In both cases, we can distinguish three frequency regimes. In the higher frequency regime, above 0.2, the 2p excitation probability exhibits a broad maximum followed by a fast monotonic decrease. The broad maximum corresponds to the one photon resonant 1s-2p transition. This maximum occurs at $\omega=0.3$ instead of 0.375. This strong shift, which is not an ac-Stark shift, results from the broad bandwidth of the pulse. Beyond the maximum, the behavior of the 2p excitation probability is easily explained by the lowest order perturbation theory in the external field. We remark that despite the extremely short  effective duration of the pulse ($< 1.5$ fs ), the 2p excitation probability doesn't depend on the value of the CEP. This contrasts with the lower frequency ($\omega < 0.02$) regime. Although the effective duration of the pulse is at least ten times longer, the 2p excitation probability is strongly sensitive to the value of the CEP $\phi$. For $\phi$ = $3\pi/2$, we observe perfect agreement between results obtained by solving numerically the TDSE and by using Eqs (\ref{eq_cs_gr_0}) and (\ref{eq_cnlm}) which give the static field limit. In this limit, the 2p-excitation probability is proportional to $\omega^2$ because, for the relatively low peak intensity and short pulse duration considered here, the imaginary part of the energy of the ground state and the relevant parabolic states stay very small. For $\phi=0$, the 2p-excitation probability is at least two orders of magnitude smaller than the same probability for $\phi=3\pi/2$. In fact, for $\phi=0$, the first order term in $\omega$ given by Eq. (\ref{eq_cs_gr_0}) is identically zero (see Eq. (\ref{eq_field_der})). It means that higher order terms, not calculated here,  become dominant.  Our TDSE results presented in Fig. \ref{fig1} show that for $\omega<0.02$ and $\phi=0$, the higher order terms are very small.  This is no longer true in the intermediate frequency regime $(0.02<\omega<0.2)$ where for both values of $\phi$, the 2p excitation probability has a highly oscillatory behavior.\\

%%%%%%%%%%% FIGURE 2
\begin{figure}[h!]
\includegraphics[width=13cm,height=8cm]{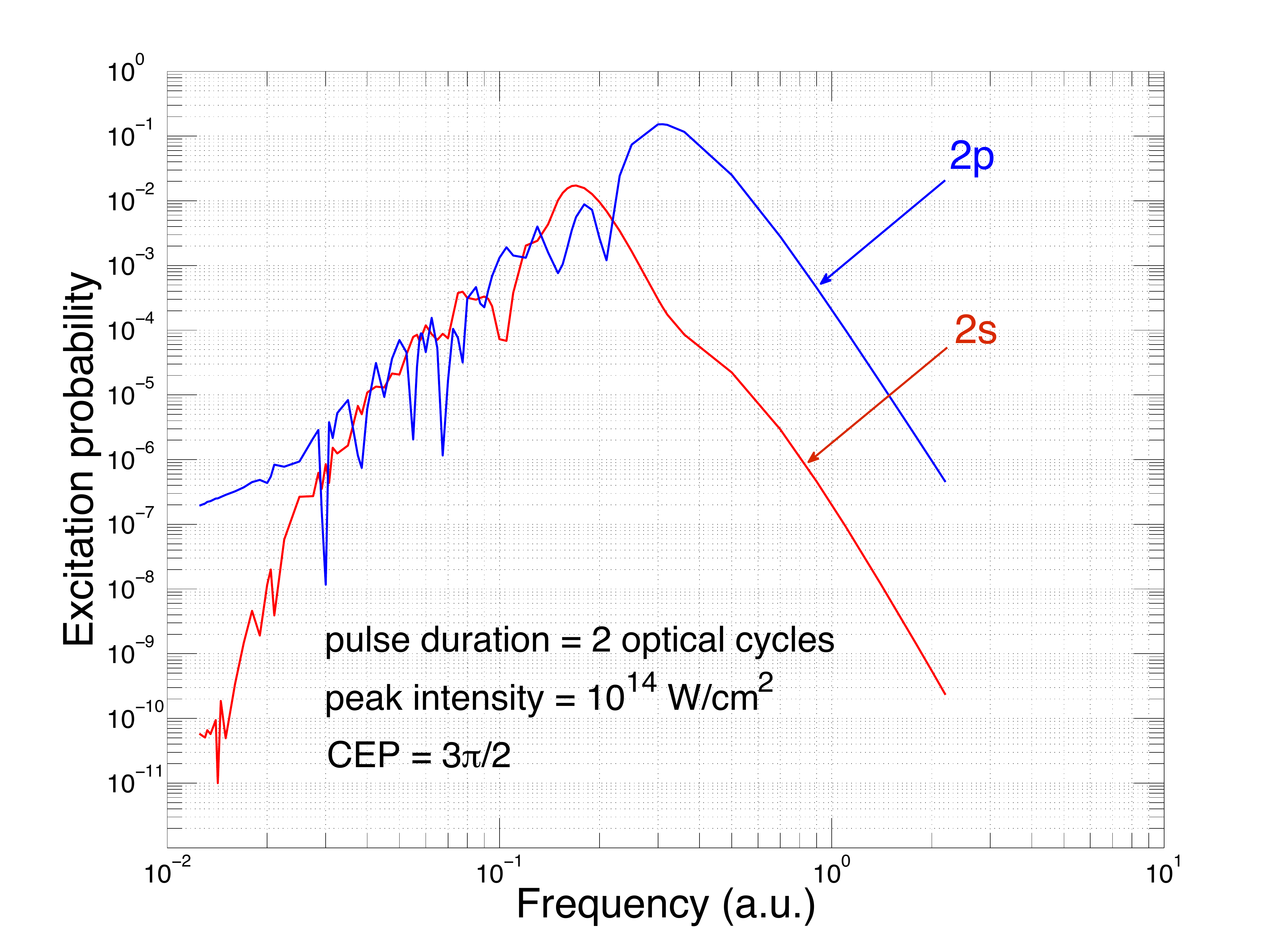}
\caption{(Color online) Probability of excitation to the 2s state (full red curve) and 2p state (full blue curve) as a function of the frequency for the same case as in Fig. \ref{fig1} and a CEP equal to $3\pi/2$.}
\label{fig2}
\end{figure}
%%%%%%%%%%%%%%%%% 
In Fig. \ref{fig2}, we compare the probability of excitation to the 2p and the 2s state for the same case as in Fig. \ref{fig1} but for a CEP of $3\pi/2$. For $\omega > 0.1$, the 2s state excitation probability exhibits a broad maximum that corresponds to a resonant two-photon 1s-2s transition. The shift observed in the 2p excitation probability is also observed for the 2s state. In the lower frequency regime ($\omega < 0.03$), the 2s state excitation probability decreases rapidly with decreasing frequencies to reach a value at $\omega=0.0125$ which is almost 4 orders of magnitude lower than the 2p state excitation probability. This difference of behavior can be explained by analyzing Eq. (\ref{eq_cnlm}). We have:\\\\
\begin{eqnarray}
\label{parspher}
\tilde{C}_{2p}(\tau_N,0)&=&\frac{1}{\sqrt{2}}\left[C_{100}(\tau_N,0)+C_{010}(\tau_N,0)\right],\\
\tilde{C}_{2s}(\tau_N,0)&=&\frac{1}{\sqrt{2}}\left[C_{100}(\tau_N,0)-C_{010}(\tau_N,0)\right].
\end{eqnarray}
It turns out that $C_{100}(\tau_N,0)\approx C_{010}(\tau_N,0)$ explaining why $\tilde{C}_{2s}(\tau_N,0)$ is very close to zero. In the regime of intermediate frequencies, the 2s state excitation probability exhibits also an oscillatory behavior but the amplitude of the oscillations is much less than in the case of the 2p state.\\
 
\begin{figure}[h!]
\includegraphics[width=13cm,height=8cm]{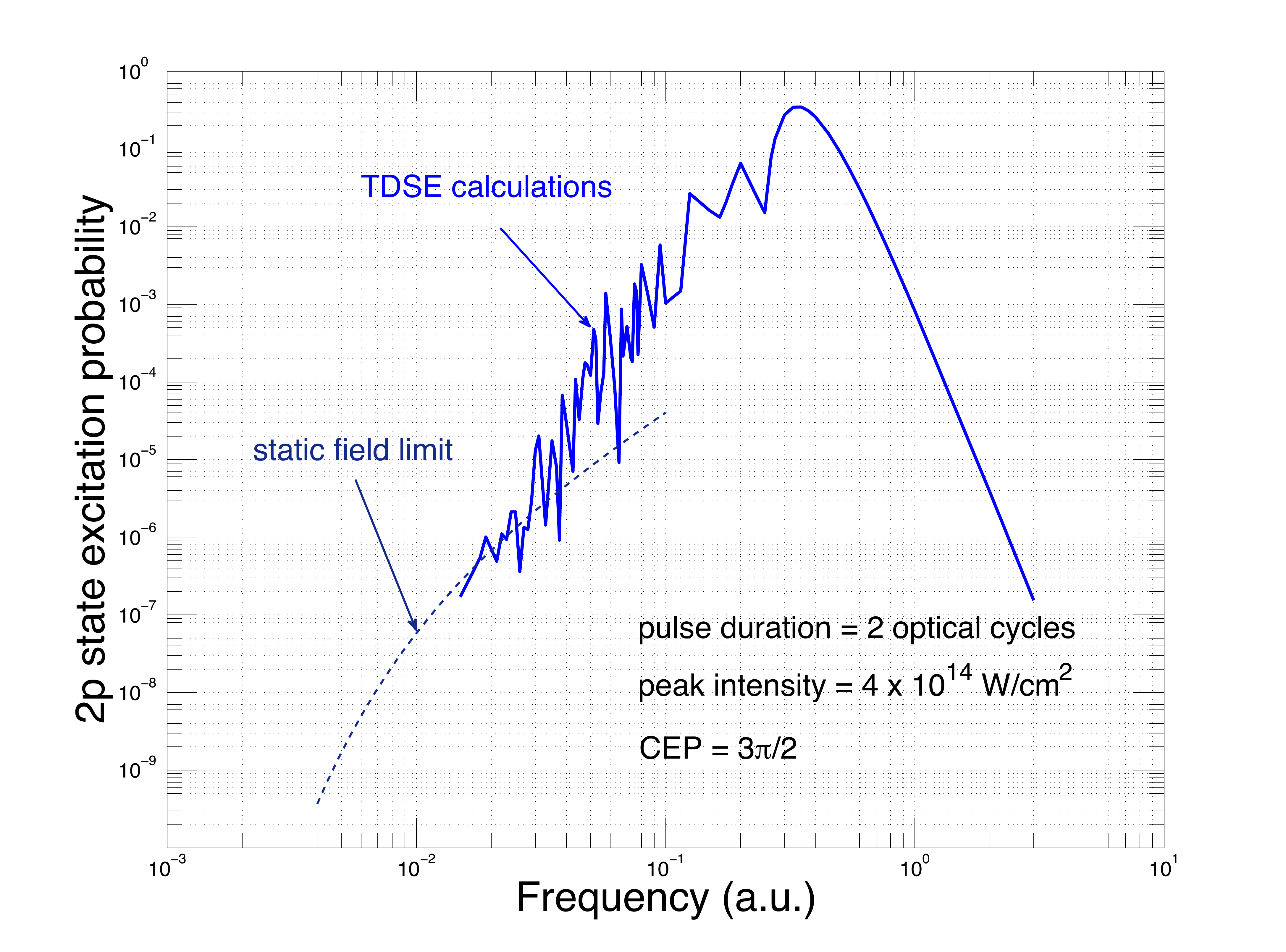}
\caption{(Color online) 2p state excitation probability as a function of the frequency in the case of the interaction of atomic hydrogen in its ground state with a 2 optical cycle sine square pulse of $4\times 10^{14}$ W/cm$^2$ peak intensity. The CEP is fixed at $3\pi/2$. Results obtained by solving numerically the TDSE (full blue line) are  compared to those (blue dashed line) obtained within the static field limit from Eq. (\ref{eq_cs_gr_0}).}
\label{fig3}
\end{figure}

In Fig. \ref{fig3}, we examine the effect of a higher peak intensity on the 2p excitation probability for the same two optical cycle sine square pulse. We consider a 
 %%%%%%%%%%% FIGURE 3
%%%%%%%%%%%%%%%%% 
peak intensity of $4\times 10^{14}$ W/cm$^2$ and compare results obtained by solving numerically the TDSE with those we get in the static field limit by using Eq. (\ref{eq_cs_gr_0}). For $\omega > 0.2$, the TDSE results show a broad maximum, which corresponds to the resonant 1s-2p transition. It is interesting to observe that the position of this maximum is closer to the expected value ($\omega=0.375$) than at $10^{14}$ W/cm$^2$. At lower frequency ($\omega < 0.02$), the 2p state excitation probability keeps oscillating by contrast to what is obtained by using Eq. (\ref{eq_cs_gr_0}) which gives the static field limit. The result obtained within this limit is no longer a straight line as is the case at a peak intensity of $10^{14}$ W/cm$^2$. This means that at low frequency, the 2p state excitation probability is no longer proportional to $\omega^2$. In fact the exponential factors (see Eq. (\ref{eq_cs_gr_0})), which are $\omega$ dependent, play a significant role at higher intensities. For intermediate frequencies, we observe again sharp oscillations as in the case of a peak intensity of $10^{14}$ W/cm$^2$.

 %%%%%%%%%%% FIGURE 4
\begin{figure}[h!]
\includegraphics[width=13cm,height=8cm]{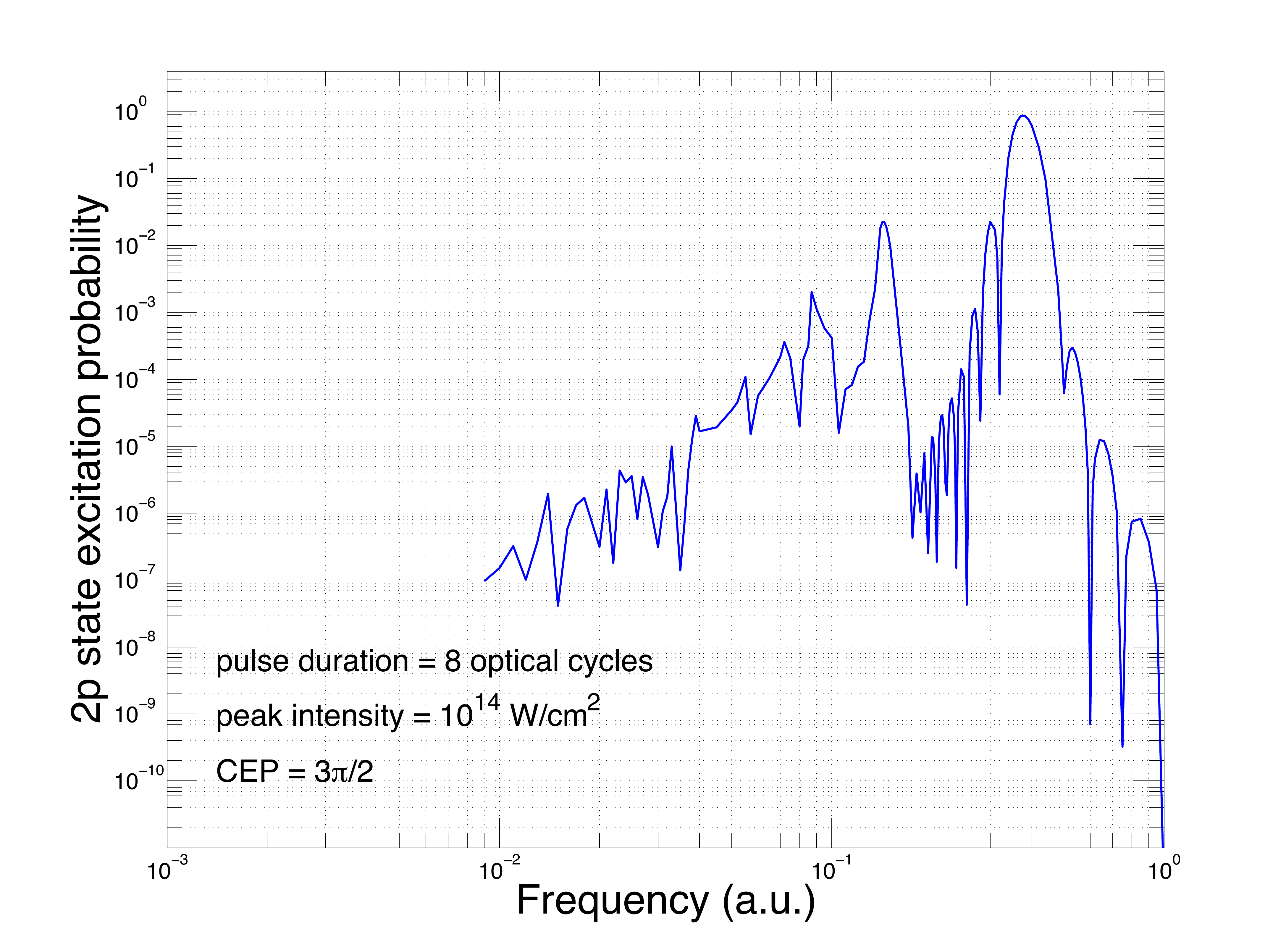}
\caption{(Color online) 2p state excitation probability as a function of the frequency in the case of the interaction of atomic hydrogen in its ground state with a 8 optical cycle sine square pulse of $10^{14}$ W/cm$^2$ peak intensity. The CEP is fixed at $3\pi/2$. The results are obtained by solving numerically the TDSE.}
\label{fig4}
\end{figure}
%%%%%%%%%%%%%%%%% 
 
In Fig. \ref{fig4}, we study the effect of increasing the number of optical cycles within the pulse on the probability of excitation to the 2p state while keeping the peak intensity equal to 10$^{14}$ W/cm$^2$. We consider an 8 optical cycle sine square pulse of $10^{14}$ W/cm$^2$ peak intensity. The CEP is fixed at $3\pi/2$. We observe a series of peaks, the first one located at a frequency $\omega=0.38$. This peak is a signature of a resonant one photon 1s-2p transition. We note that, for 8 optical cycles, there is a small ac-Stark shift, positive as expected, and no significant shift resulting from the spectral bandwidth of the pulse. The sidebands on both sides of this peak are reminiscent of the Fourier transform of the sine square pulse. These sidebands are not resolved in the case of the much shorter 2 optical cycle pulse.  For frequencies between 0.04 and 0.2 several resonant peaks are present which, tentatively, we associate to resonant multiphoton transitions. However, identifying the exact position of such peaks would require a full Floquet based treatment, out of reach at the present moment for such laser parameters. At lower frequencies, our results exhibit a complex behavior which doesn't  match the behavior expected in the static field limit. The 2p excitation probability within this limit (not shown in the Fig. \ref{fig4}) is found two orders of magnitude below the TDSE results, meaning that this limit should be reached at much lower frequencies. In this frequency range, higher order terms  in the iterative scheme developed here, are expected to play a role. However, their calculation is very difficult to perform since all complex energies have to be generated as a function of the electric field within the pulse.\\
 
\subsection{Total excitation and ionization probability}

%%%%%%%%%%% FIGURE 5
\begin{figure}[h!]
\includegraphics[width=13cm,height=8cm]{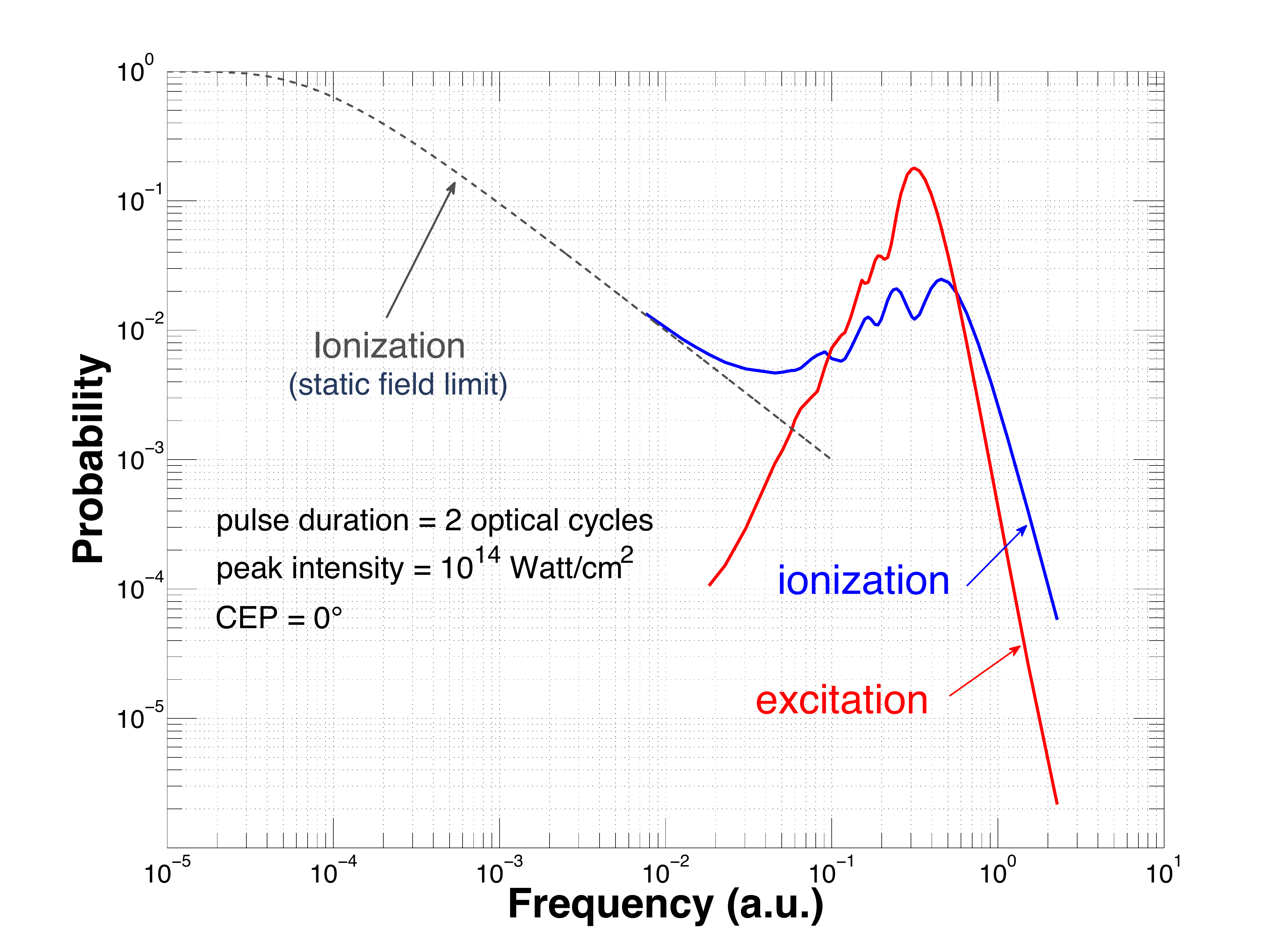}
\caption{(Color online) Total excitation (full red line) and ionization (full blue line) probability as a function of the frequency in the case of the interaction of atomic hydrogen in its ground state with a 2 optical cycle sine square pulse of $10^{14}$ W/cm$^2$ peak intensity. The CEP $\phi=0$. The results are obtained by solving numerically the TDSE. The dashed line represents the ionization probability in the static field limit.}
\label{fig5}
\end{figure}
%%%%%%%%%%%%%%%%% 
%%%%%%%%%%% FIGURE 6
\begin{figure}[h!]
\includegraphics[width=13cm,height=8cm]{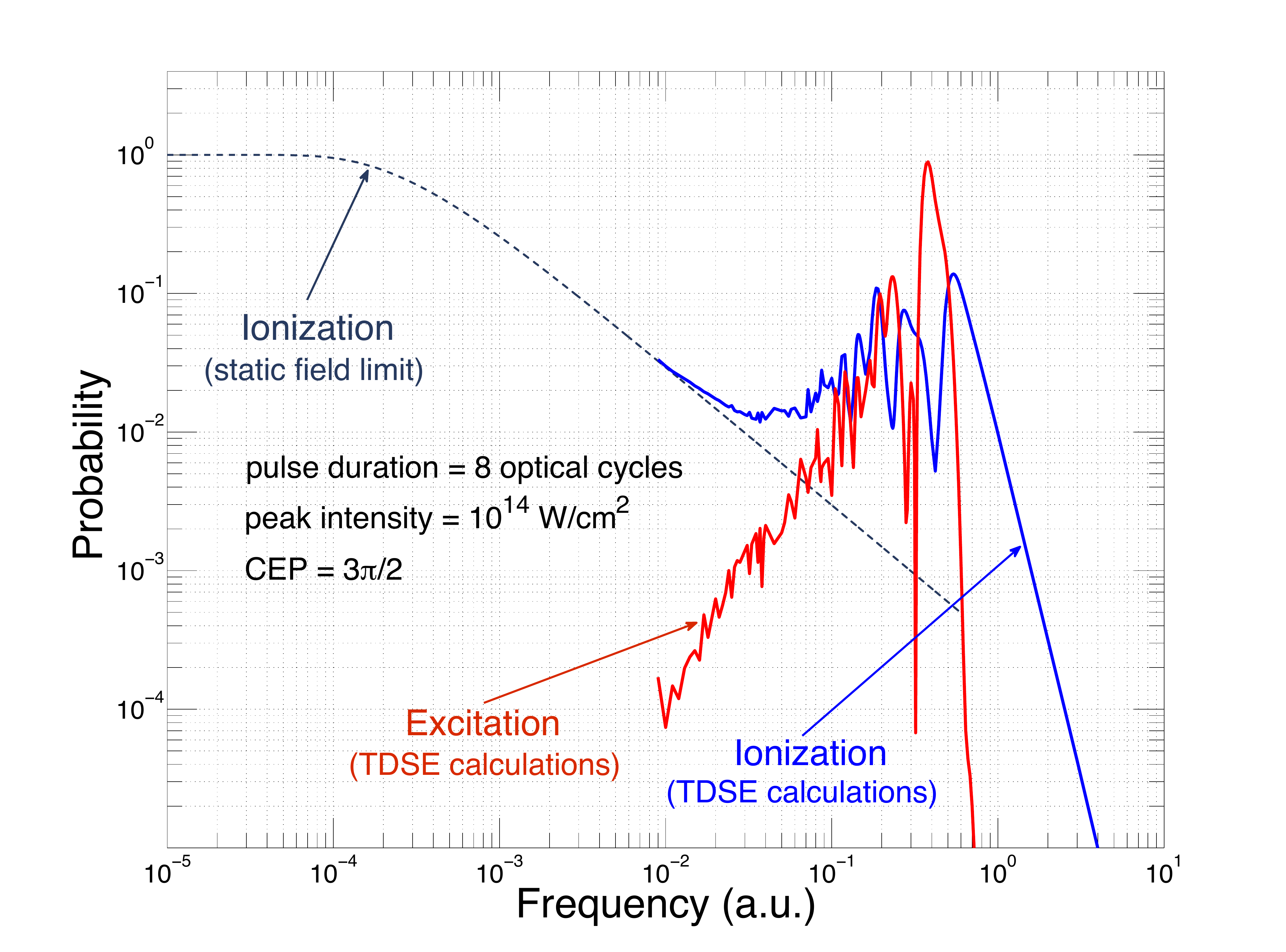}
\caption{(Color online) Total excitation (full red line) and ionization (full blue line) probability as a function of the frequency in the case of the interaction of atomic hydrogen in its ground state with a 8 optical cycle sine square pulse of $10^{14}$ W/cm$^2$ peak intensity. The CEP $\phi=3\pi/2$. The results are obtained by solving numerically the TDSE. The dashed line represents the ionization probability in the static field limit.}
\label{fig6}
\end{figure}
%%%%%%%%%%%%%%%%% 
In this Section, we examine the effect of the number of cycles for the sine square pulse on the total excitation probability as a function of the frequency and compare it to the ionization probability. In Fig. \ref{fig5}, we consider a two optical cycle pulse of $10^{14}$ W/cm$^2$ peak intensity with a CEP $\phi=0$. The total excitation probability exhibits a broad resonance with the maximum at $\omega=0.315$. By comparing with Fig. \ref{fig1}, we see that this resonance results mainly from the excitation to the 2p state. However, contrary to the 2p excitation probability, the low frequency behavior of the total excitation probability is smooth and up to five orders of magnitude higher. This clearly indicates that the contribution of other excited states becomes significant. In fact, a more detailed analysis shows that there is a cumulative contribution of many excited states including high-lying Rydberg states \cite{piraux17}. The smoothness of the total excitation probability at low frequency is expected due to the shortness of the pulse used. For frequencies around the main excitation resonance, the ionization probability shows oscillations which are approximately in phase opposition with the slight oscillations observed in the total excitation probability. In fact, the ionization maxima result from channel closings. For decreasing frequencies, the ionization probability presents a minimum at $\omega=0.05$ (see \cite{Dimi} before tending smoothly to the static field limit evaluated at the zeroth order from Eq. (\ref{eq_w00}).\\

In Fig. \ref{fig6}, we consider an 8 optical cycle pulse of $10^{14}$ W/cm$^2$ peak intensity with a CEP $\phi=3\pi/2$. The total excitation probability exhibits a sharp maximum at $\omega=0.38$, which results predominantly from the resonant one photon excitation to the 2p state. However, at lower frequency, the total excitation probability shows a rich structure, as a consequence of the longer pulse duration. In addition and similarly to the case of the two-optical cycle pulse, the total excitation is now about three orders of magnitude higher than the 2p excitation probability (see Fig. \ref{fig4}). This is again the result of the cumulative contributions of many excited states including high lying Rydberg states. Fig. \ref{fig6} shows also the TDSE calculations of the ionization probability where we observe many oscillations clearly in phase opposition with the oscillations of the total excitation probability. As the frequency decreases, this ionization probability reaches its static field limit. 
 
\section{CONCLUSIONS}

In this paper, we considered the interaction of atomic hydrogen in its ground state with a laser pulse. The main objective was to derive an analytical expression for the probability of excitation to a given bound state for increasingly small frequencies while keeping constant the number of optical cycles within the pulse. This is what we referred to as the static field limit. The main idea behind this derivation is to transform the total hamiltonian by using a complex scaling of the radial coordinate with an angle between 0 and $\pi/3$ for which the spectrum is purely discrete. In this case, the complex eigenvalues correspond to quasi-bound states, which are the analytical continuation of the bound states of atomic  hydrogen when the electric field is turned on. The resulting formula depends on three parameters, the peak intensity, the number of cycles within the pulse and the value of the CEP.\\

To analyze the validity of this formula, we calculated the probability of excitation to a given state with this method and compared the results with those obtained by solving numerically the TDSE, for different values of the three pertinent parameters. We studied the probability of excitation to the 2p and 2s state and concluded that the analytical formula agrees with the TDSE results for low peak intensities and short pulses of few optical cycles and for CEP different from zero. For CEP=0, higher order terms in the perturbative expansion of the excitation probability amplitude have to be taken into account. These high order terms involve virtual dipolar transitions to many parabolic states in presence of the slowly varying electric field.\\

A clear assessment of the domain of validity of the analytical formula is very difficult. It depends on both the pulse duration and the peak intensity. For long pulse durations, the peak intensity should be of the order or less than the intensity at which the first quasi crossings of the complex energies occur since beyond this quasi crossings, the parabolic states lose their identity. However, for short pulse durations, the presence of these quasi crossings at low field intensity plays a minor role because the imaginary part of the complex energies (which is related to the width of the corresponding  parabolic states) is very small so that the absolute value of the exponential terms in the analytical formula is about one, leading to a $\omega^2$ behavior for the excitation probability. For longer pulse durations, this is no longer true. In that case, the electric field strength at which the first quasi crossing occurs fixes the peak intensity at which the analytical formula is valid. This peak intensity strongly depends, as expected, on the principal quantum number of the excited state one considers. As a matter of fact, our analytical result for the low-frequency excitation probability to a given excited state should be valid provided that the field frequency is  much smaller than any frequency corresponding to a transition from the final excited state to neighboring states. If it is not the case, the final state could be resonantly coupled to other neighboring states. This situation will be investigated in a forthcoming publication. Another way to view this, is to say that if the field frequency is higher than some transition frequencies, we are no longer in the adiabatic regime.\\ 

Finally, we discussed results obtained by solving numerically the TDSE, for the total excitation and ionization probability as a function of frequency for laser pulses of 2 and 8 optical cycles and a peak intensity equal to 10$^{14}$ W/cm$^2$. For intermediate va\-lues of the frequency, the excitation and ionization probabilities show clear out of phase oscillations. The maxima of the ionization probability correspond to channel closings showing that the mechanism of both excitation and ionization involve multiphoton transitions. For lower frequencies, and a pulse of  8 optical cycles, the excitation probability exhibits fast oscillations. The origin of these fast oscillations is not clear and requires further investigations.\\

\section{Acknowledgements}
The authors thank Prof. N.N.  Nefedov and V.Y. Popov for very valuable discussions on the mathematical aspects of the present work. A.G. was "aspirant au Fonds de la Recherche Scientifique (F.R.S-FNRS)"; he acknowledges the F.R.S-FNRS for having given him the opportunity to do and to complete his PhD thesis.  Y.P.  thanks the Universit\'e Catholique de Louvain (UCL) for financially supporting several stays at the Institute of Condensed Matter and Nanosciences of the UCL. F.M.F and P.F.O'M gratefully acknowledge the European network COST (Cooperation in Science and Technology) through the Action CM1204 "XUV/X-ray light and fast ions for ultrafast chemistry" (XLIC) for financing several short term scientific missions at UCL. The present research benefited from computational resources made available on the Tier-1 supercomputer of the F\'ed\'eration Wallonie-Bruxelles funded by the R\'egion Wallonne under the grant n$^o$1117545 as well as on the supercomputer Lomonosov from Moscow State University and on the supercomputing facilities of the UCL and the Consortium des Equipements de Calcul Intensif (CECI) en F\'ed\'eration Wallonie-Bruxelles funded by the F.R.S.-FNRS under the convention 2.5020.11. Y.P. is grateful to Russian Foundation for Basic Research (RFBR) for the financial support under the grant N$^0$ 16-02-00049-a.

\section*{Appendix}

In configuration space, Eqs (12) and (13) take the form :
\begin{equation}
[\varepsilon_j(\tau)-H(\tau,\vec{r} e^{\theta})]\tilde\psi_j(\tau,\vec{r} e^{\theta})=0, 
\end{equation} 
\begin{equation}
\;\;\;[\varepsilon_j(\tau)-H(\tau,\vec{r} e^{\theta})]\tilde\psi_j^*(\tau,\vec{r} e^{\theta^*})=0,
\end{equation}
and the normalization condition (\ref{eq_norm}) becomes,
\begin{equation}
\int\mathrm{d}^3r\ \tilde\psi_i(\tau,\vec{r} e^{\theta})\tilde\psi_j(\tau,\vec{r} e^{\theta})=\delta_{ij}.
\end{equation}
In particular, for $i=j$, we have
\begin{equation}
\int\mathrm{d}^3r\ \tilde\psi_j^2(\tau,\vec{r} e^{\theta})=1.
\end{equation}
By differentiating this expression with respect to $\tau$, we obtain
\begin{equation}
\langle\psi_j(\tau,\theta^*)|\frac{\partial\psi_{j}(\tau,\theta)}{\partial\tau}\rangle=0.
\end{equation}
A similar relation can also be obtained for degenerate states. The proof is based
on the Kato-Rellich perturbation theory \cite{Kato,Benassi}. Let us consider the 
projector $P_j(\tau+d\tau,\theta)$ associated with the hamiltonian $H(\tau+d\tau,\theta)\approx
H(\tau,\theta)+(\partial/\partial\tau) V(\tau,\theta)d\tau + ...$ where $d\tau$ is supposed to be 
very small. According to \cite{Kato,Benassi}, we write
\begin{eqnarray}
P_j(\tau+d\tau,\theta)&\approx & P_j(\tau,\theta)+G(\varepsilon_j(\tau),\tau,\theta)\frac{\partial V(\tau,\theta)}{\partial\tau}P_j(\tau,\theta)\mathrm{d}\tau \nonumber\\
&+&P_j(\tau,\theta)\frac{\partial V(\tau,\theta)}{\partial\tau}G(\varepsilon_j(\tau),\tau,\theta)\mathrm{d}\tau + ...\;\; ,       
\end{eqnarray}
where the Green's function $G(z,\tau,\theta)$ is given by
\begin{eqnarray}
G(z,\tau,\theta)&=& \left(z-H(\tau,\theta)\right)^{-1}\left(I-P_j(\tau,\theta)\right)\nonumber\\
&=&\left(z-H(\tau,\theta)\right)^{-1}\left(I-\sum_{s=1}^r|\psi_{j_s}(\tau,\theta)><\psi_{j_s}(\tau,\theta^*)|\right),
\end{eqnarray}
with $r$ being the degree of degeneracy of the state $j$. We can rewrite the Schr\"odinger equation as follows:
\begin{equation}
H(\tau+\mathrm{d}\tau,\theta)P_j(\tau+\mathrm{d}\tau,\theta)|\psi_{j_s}(\tau,\theta)\rangle=\varepsilon_j(\tau)P_j(\tau+\mathrm{d}\tau,\theta)|\psi_{j_s}(\tau,\theta)\rangle. 
\end{equation}
As a result, we have:
\begin{eqnarray}
|\psi_{j_s}(\tau+\mathrm{d}\tau,\theta)\rangle&\approx&
P_j(\tau+\mathrm{d}\tau,\theta)|\psi_{j_s}(\tau,\theta)\rangle + ... \nonumber
\end{eqnarray}
\begin{equation}
\;\;\;\;\;=P_j(\tau,\theta)|\psi_{j_s}(\tau,\theta)\rangle +G(\varepsilon_j(\tau),\tau,\theta)\frac{\partial V(\tau,\theta)}{\partial\tau}P_j(\tau,\theta)|\psi_{j_s}(\tau,\theta)\rangle \mathrm{d}\tau + ...\;\;\; . 
\end{equation}
From this relation, it follows that
\begin{equation}
\langle\psi_{j_l}(\tau,\theta^*)|\frac{\partial\psi_{j_s}(\tau,\theta)}{\partial \tau}\rangle=\langle\psi_{j_l}(\tau,\theta^*)|G(\varepsilon_j(\tau),\tau,\theta)\frac{\partial
V(\tau,\theta)}{\partial\tau}|\psi_{j_s}(\tau,\theta)\rangle=0. 
\end{equation}
This result is also given in Part 2, Ch.II, \S 6 of \cite{Fock}.\\

\printbibliography

\end{document}